\documentclass[twocolumn,showpacs,preprintnumbers,amsmath,amssymb,superscriptaddress]{revtex4}
\usepackage{graphicx}
\usepackage{bm}
\usepackage{color}
\usepackage{amsmath}
\usepackage{amsfonts}
\usepackage{amssymb}
\usepackage{epstopdf}

\usepackage{ulem}
\usepackage{array}

\newcolumntype{L}[1]{>{\raggedright\let\newline\\\arraybackslash\hspace{0pt}}m{#1}}
\newcolumntype{C}[1]{>{\centering\let\newline\\\arraybackslash\hspace{0pt}}m{#1}}
\newcommand{\bra}[1]{\langle #1|}
\newcommand{\ket}[1]{|#1\rangle}

\begin{document}

\preprint{APS/123-QED}

\title{Quasi-two-dimensional Bose-Einstein condensation of spin triplets
in dimerized quantum magnet Ba$_2$CuSi$_2$O$_6$Cl$_2$}
% Force line breaks with \\

%\title{}

\author{Makiko Okada}
\author{Hidekazu Tanaka}
\email{tanaka@lee.phys.titech.ac.jp}
\author{Nobuyuki Kurita}
\affiliation{Department of Physics, Tokyo Institute of Technology,
  Meguro-ku, Tokyo 152-8551, Japan}

\author{Kohei Johmoto} 
\author{Hidehiro Uekusa}
\affiliation{Department of Chemistry, Tokyo Institute of Technology,
  Meguro-ku, Tokyo 152-8551, Japan}

\author{Atsushi Miyake}
\author{Masashi Tokunaga}
\affiliation{Institute for Solid State Physics, University of Tokyo, Kashiwa,
  Chiba 277-8581, Japan}

\author{Satoshi Nishimoto}
\affiliation{Institute for Theoretical Solid State Physics, IFW
  Dresden, 01171 Dresden, Germany}
\affiliation{Department of Physics, Technical University Dresden, 01069 Dresden, Germany}

\author{Masaaki Nakamura}
\affiliation{Department of Physics, Ehime University,
Bunkyo-cho 2-5, Matsuyama, Ehime 790-8577, Japan}

\author{Marcelo Jaime}
 
\affiliation{National High Magnetic Field Laboratory,
    and Materials, Physics and Applications Division,\\Los Alamos
    National Laboratory, Los Alamos, NM 87545, USA}

\author{Guillaume Radtke}
\affiliation{Institut de Min\'eralogie, de Physique des
    Mat\'eriaux et de Cosmochimie, Sorbonne Universit\'es,
    Universit\'e Pierre et Marie Curie - Paris 6, UMR CNRS 7590,
    Museum National d'Histoire Naturelle, IRD UMR 206, 75005
    Paris, France}

\author{Andr\'es Sa\'ul}
\affiliation{Aix-Marseille University, Centre
    Interdisciplinaire de Nanoscience de Marseille-CNRS UMR 7325\\
    Campus de Luminy, 13288 Marseille cedex 9, France}

\date{\today}% It is always \today, today,

\begin{abstract}
  
  We synthesized single crystals of composition
  Ba$_2$CuSi$_2$O$_6$Cl$_2$ and investigated its quantum magnetic
  properties. The crystal structure is closely related to that of the
  quasi-two-dimensional (2D) dimerized magnet BaCuSi$_2$O$_6$ also
  known as Han purple. Ba$_2$CuSi$_2$O$_6$Cl$_2$ has a singlet ground
  state with an excitation gap of ${\Delta}/k_{\rm B}\,{=}\,20.8$ K.
  The magnetization curves for two different field directions almost
  perfectly coincide when normalized by the $g$-factor except for a
  small jump anomaly for a magnetic field perpendicular to the $c$
  axis. The magnetization curve with a nonlinear slope above the
  critical field is in excellent agreement with exact-diagonalization
  calculations based on a 2D coupled spin-dimer model. Individual
  exchange constants are also evaluated using density functional
  theory (DFT). The DFT results demonstrate a 2D exchange network and
  weak frustration between interdimer exchange interactions, supported
  by weak spin-lattice coupling implied from our magnetostriction
  data. The magnetic-field-induced spin ordering in
  Ba$_2$CuSi$_2$O$_6$Cl$_2$ is described as the quasi-2D Bose-Einstein
  condensation of triplets.
\end{abstract}

%http://www.aip.org/pacs/

\maketitle

\section{Introduction\label{intro}}
Bose-Einstein condensation (BEC) is a fascinating macroscopic quantum
phenomenon characteristic of system of bosons~\cite{Bose,Einstein}.
BEC occurs not only in true bosonic
particles~\cite{London,Anderson,Davis} but also in bosonic
quasiparticles~\cite{Deng,Byrnes}. Dimerized quantum magnets (DQMs)
provide a stage to embody the quantum physics of interacting lattice
bosons~\cite{Rice,Sachdev2}. In a DQM, two spins are dimerized by a
strong antiferromagnetic (AFM) exchange interaction and these dimers
are coupled via weak exchange interactions. DQMs usually have a gapped
singlet ground state and exhibit
magnetic-field-induced~\cite{Oosawa1,Oosawa2,Rueegg,Jaime,Sebastian,Yamada}
and pressure-induced~\cite{Oosawa3,goto_t,O_n,Rueegg2,Rueegg3,goto_m}
quantum phase transitions (QPTs) to an ordered state. These QPTs can
be described as the BEC of the $S_z\,{=}\,{\pm}1$ components of the
spin triplets called
triplons~\cite{Nikuni,Matsumoto,Nohadani2,Giamarchi,Zapf}.
In pressure-induced BEC, the triply degenerate triplons are
reconstructed at the quantum critical point (QCP) into one amplitude
mode~\cite{Matsumoto}, which can be interpreted as the Higgs mode, and
two phase modes, which correspond to the Nambu-Goldstone
modes~\cite{Pekker}. In magnetic-field-induced BEC of triplons with
$S_z\,{=}\,{+}\,1$, the particle quantities of chemical potential
$\mu$, triplon density $n$ and compressibility
${\kappa}\,{=}\,{\partial}n/{\partial}{\mu}$ correspond to the
magnetic quantities of the external field $H$, total magnetization $M$
and magnetic susceptibility ${\chi}\,{=}\,{\partial}M/{\partial}H$,
respectively, which are easily measured. A feature of triplon BEC is
that the total number of triplons is controllable by tuning the
magnetic field or pressure, in contrast to the BEC of true particles.

The magnetic-field-induced BEC of triplons in 3D DQMs has been studied
experimentally for many
systems~\cite{Oosawa1,Oosawa2,Rueegg,Jaime,Sebastian,Yamada},
and the results observed were quantitatively or semiquantitatively
described by the triplon BEC
theory~\cite{Nikuni,Matsumoto,Yamada}. On the other
hand, in 2D systems, long-range ordering is suppressed by large
quantum and thermal fluctuations, and topological ordering such as the
Berezinskii-Kosterlitz-Thouless (BKT)
transition~\cite{Berezinskii,KT1,KT2} and unusual quantum critical
behavior are expected to occur. Recently, the BKT transition has been
reported for the 2D DQM
C$_{36}$H$_{48}$Cu$_2$F$_6$N$_8$O$_{12}$S$_2$~\cite{BKT}. However, the
understanding of 2D systems of interacting triplons is insufficient.
For a detailed study of the systems, a new 2D-like DQM, for which
sizable single crystals are obtainable, is necessary. In this paper,
we show that Ba$_2$CuSi$_2$O$_6$Cl$_2$, synthesized in this work, is
close to an ideal 2D dimerized isotropic quantum magnet, which
exhibits field-induced magnetic order in laboratory-accessible
critical fields.

\section{Experimental details\label{experimental}}
To synthesize single crystals of Ba$_2$CuSi$_2$O$_6$Cl$_2$, we first
prepared Ba$_2$CuTeO$_6$ powder by a solid-state reaction.
%~\cite{BaCuTeO6}. 
A mixture of Ba$_2$CuTeO$_6$ and BaCl$_2$ in a molar ratio of
$1\,{:}\,10$ was vacuum-sealed in a quartz tube. The temperature at
the center of the horizontal tube furnace was lowered from 1100 to
800$^{\circ}$C over 10 days. Plate-shaped blue single crystals with a
maximum size of $3\,{\times}\,3{\times}\,1$ mm$^3$ were obtained.
These crystals were found to be Ba$_2$CuSi$_2$O$_6$Cl$_2$ from X-ray
diffraction, where the silicon was provided by the quartz tube. The
wide plane of the crystals was confirmed to be the crystallographic
$ab$ plane by X-ray diffraction. We found that the quartz tube
frequently exploded during cooling to room temperature after the
crystallization process from 1100 to 800$^{\circ}$C.  To avoid
hazardous conditions and damage to the furnace, a cylindrical nichrome
protector was inserted in the furnace core tube.
%Details of the structural analysis carried out on Ba$_2$CuSi$_2$O$_6$Cl$_2$ (BCuSOC) are described in the Supplemental Materials~\cite{Supplement}.

The magnetic susceptibility of Ba$_2$CuSi$_2$O$_6$Cl$_2$ single
crystal was measured in the temperature range of $1.8\,{-}\,300$ K
using a SQUID magnetometer (Quantum Design MPMS XL). The specific heat
of Ba$_2$CuSi$_2$O$_6$Cl$_2$ was measured in the temperature range of
$0.5\,{-}\,300$ K using a physical property measurement system
(Quantum Design PPMS) by the relaxation method. High-field
magnetization measurement in a magnetic field of up to 45 T was
performed at 4.2 and 1.4 K using an induction method with a multilayer
pulse magnet at the Institute for Solid State Physics, University of
Tokyo.

Magnetostriction of Ba$_2$CuSi$_2$O$_6$Cl$_2$ was measured at National High Magnetic Field Laboratory, Los Alamos National Laboratory.
Variations in the sample length $L$ as a function of the temperature
and/or magnetic field, ${\varepsilon}(H)={\Delta}L/L (H)=[L(H, T)-L(H_0, T_0)]/L(H_0, T_0)$, were measured using a fiber Bragg
grating (FBG) technique~\cite{daou,jaime} consisting of recording
spectral information of the light reflected by a 0.5-mm-long Bragg
grating inscribed in the core of a 125\,$\mu$m telecom optical fiber.
The FBG section of the fiber is attached to the sample to be studied, and
changes in the grating spacing are driven by changes in the sample
dimensions when the temperature or magnetic field is changed. This technique 
has a demonstrated resolution of $\Delta L/L \simeq 10^{-7}$ that is virtually 
immune to the electromagnetic noise characteristic of pulsed magnetic fields but is somewhat sensitive to mechanical vibrations. Capacitor-bank-driven pulsed magnets were used to produce 25\,ms magnetic field pulses up to 60\,T.

\begin{table}[thb]
\caption{Crystal data for Ba$_2$CuSi$_2$O$_6$Cl$_2$.}
\label{table:1}
\begin{ruledtabular}
\begin{tabular}{cccc}
& Chemical formula & Ba$_2$CuSi$_2$O$_6$Cl$_2$ &  \\
& Space group & $Cmca$ &  \\
& $a$ ($\rm{\AA}$) & 13.8917(12) &  \\
& $b$ ($\rm{\AA}$) & 13.8563(11) &  \\
& $c$ ($\rm{\AA}$) & 119.6035(15) &  \\
& $V$ ($\rm{\AA}^3$) & 3773.4(5) &  \\
& $Z$ & 16 &  \\
%& No. of observed reflections & 2163 ($I\,{>}\,2{\sigma}(I)$) & \\
& $R;\ wR$ &  0.0773;\ 0.1638 & 
\end{tabular}
\end{ruledtabular}
\end{table}

\begin{table}[ht]
\caption{Fractional atomic coordinates (${\times}\,10^4$) and equivalent isotropic displacement parameters ($\rm{\AA}^2{\times}\,10^3$) for Ba$_2$CuSi$_2$O$_6$Cl$_2$.}
\label{table:2}
\begin{ruledtabular}
\begin{tabular}{rrrrr}
Atom  &  $x$\hspace{7mm}    & $y$\hspace{7mm}   & $z$\hspace{5mm}  & $U_{\rm eq}$\hspace{1mm}   \\ \hline
Ba(1) & 2715(1) & 6230(1) & 1093(1) & 25(1)\\
Ba(2) & 0 & 3917(1) & 1045(1) & 24(1)\\
Ba(3) & 5000 & 3535(1) & 1046(1) & 27(1)\\
Cl(1)    & 2512(5) & 3562(3) & 429(3) & 51(2)\\
Cl(2)    & 0 & 1390(9) & 569(4) & 68(3)\\
Cl(3)    & 5000 & 1244(6) & 585(5) & 74(3)\\
Cu(1) & 2503(1) & 3783(2) & 1761(1) & 17(1)\\
O(1) & 1747(8) & 4919(9)         & 1819(7) & 24(3)\\
O(2) & 3693(9) & 4495(8)         & 1820(7) & 26(3)\\
O(3) & 1231(8) & 6241(8) & 2728(6) & 21(2)\\
O(4) & 0 & 4941(12) & 2251(9)    & 22(4)\\
O(5) & 3278(8) & 2565(8) & 1827(6) & 20(2)\\
O(6) & 0 & 2532(11)      & 2750(8) & 19(3)\\
O(7) & 1334(8) & 2960(8) & 1822(6) & 20(2)\\
Si(1) & 1111(4)  & 5122(3) & 2495(3) & 21(1)\\
Si(2) & 1109(4) & 2376(3) & 2499(2) & 15(1) 
\end{tabular}
\end{ruledtabular}
\end{table}

%{\color{red}
\section{Computational details}

Band structure and total energy calculations were performed with the
\textsc{Quantum Espresso}~\cite{QuantumEspresso} suite of codes
based on density functional theory (DFT) and using the pseudopotential
plane-wave method. The calculations were performed using ultrasoft
pseudopotentials~\cite{Garrity14} with plane-wave and charge
density cutoffs of 60 Ry and 400 Ry, respectively, and a $4\,{\times}\,4\,{\times}\,2$
Monkhorst-Pack~\cite{Monkhorst1976} grid for the first Brillouin
zone sampling of the 104-atom base-centered orthorhombic unit cell of
Ba$_2$CuSi$_2$O$_6$Cl$_2$.
  
Exchange and correlation were taken into account using the
generalized gradient approximation of Perdew, Burke and
Ernzerhof~\cite{Perdew96} (GGA-PBE) augmented by a Hubbard $U$ term
to improve the treatment of strongly correlated Cu-3$d$ electrons.  A
value of $U_{\mathrm{scf}}\,{=}\,10.4$~eV for the effective
self-consistent Hubbard term was determined for the 104-atom unit cell following the approach
described in Refs. \cite{Cococcioni05}
and~\cite{Kulik06}, using the
experimental structure determined at 300~K and a ferromagnetic
order.

\begin{figure*}[ht!]
  \centering \includegraphics[width=17.0cm,clip]{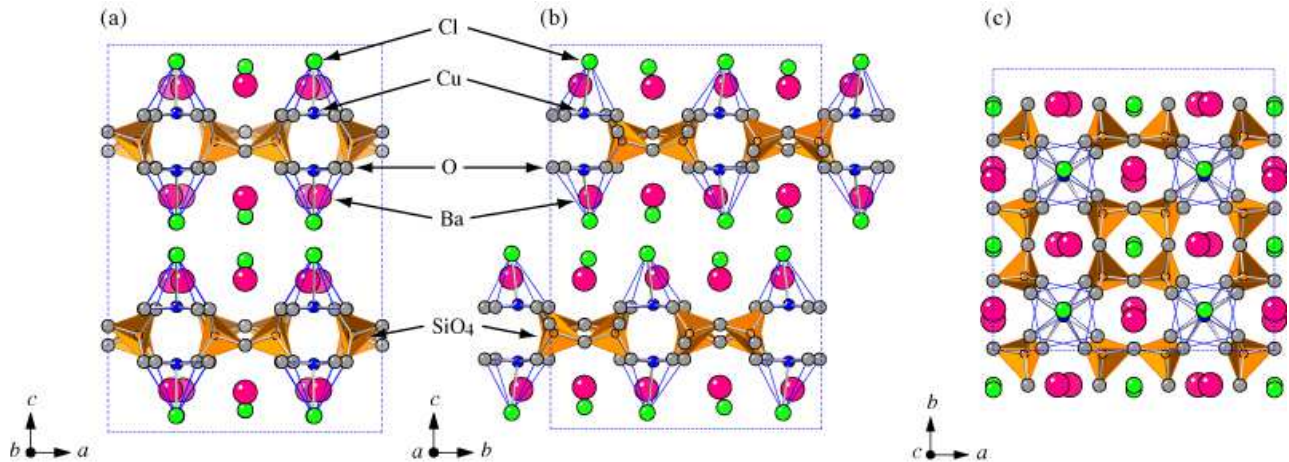}
\caption{(Color online) Crystal structure of
  Ba$_2$CuSi$_2$O$_6$Cl$_2$ viewed along the (a) $b$ axis, (b) $a$
  axis and (c) $c$ axis.}
\label{fig:Cu_structure}
\end{figure*}
  
Maximally localized Wannier function~\cite{Marzari97} (MLWF) 
interpolation of the band structure was performed using
\textsc{Wannier90}~\cite{Wannier90}, which allows the extraction
of effective hopping integrals between magnetic orbitals.
  
The calculation of magnetic couplings was carried out within the
broken symmetry formalism, \textit{i.e.}, by mapping total energies
corresponding to various collinear spin arrangements within a
supercell~\cite{footnote1} onto the Heisenberg Hamiltonian
\begin{equation}
\label{Heisenberg}
{\cal H} = {\cal H}_0 + \sum_{\langle i,j\rangle} J_{ij} \; {\bm{S}}_{i} \cdot
{\bm{S}}_{j}, 
\end{equation}
where ${\cal H}_0$ is the spin-independent part of the Hamiltonian,
$J_{ij}$ are the magnetic couplings to be determined, and
${\bm{S}}_{i}$ and ${\bm{S}}_{j}$ are, in our case,
the $S,{=}\,1/2$ spin operators localized on Cu$^{2+}$ ions located at sites
$i$ and $j$ respectively. It is straightforward to show that the
expectation value of the Hamiltonian~(\ref{Heisenberg}) on a DFT state
$\ket{\alpha}$ (obtained by preparing the initial electron density
according to a particular collinear spin arrangement in the supercell
and performing a self-consistent calculation until convergence) can be
simply written in the form of the Ising Hamiltonian~\cite{Radtke10}
\begin{equation}
\label{Ising}
\epsilon_{\alpha}^{\rm DFT} = \bra{\alpha} {\cal H} \ket{\alpha} =\epsilon_0
+ \frac{1}{4} \sum_{\langle i,j\rangle} J_{ij} \sigma_i \sigma_j
\end{equation}
with $\sigma_i\,{=}\,{\pm} 1$. In strongly localized systems, such as $3d$
transition-metal oxides, Eq.~(\ref{Ising}) can be employed to model
large sets of spin configurations, and a least-squares minimization of
the difference between the DFT and Ising relative energies can be applied
to obtain a numerical evaluation of the
couplings~\cite{Saul11,Saul14}. 
%}

\section{Crystal structure\label{structure}}

We performed a structural analysis at room temperature using a RIGAKU
R-AXIS RAPID three-circle X-ray diffractometer equipped with an
imaging plate area detector.  Monochromatic Mo-K$\alpha$ radiation
with a wavelength of ${\lambda}\,{=}\,0.71075$\,\rm{\AA} was used as
the X-ray source. Data integration and global-cell refinements were
performed using data in the range of
$3.119^{\circ}\,{<}\,{\theta}\,{<}\,27.485^{\circ}$, and absorption
correction based on face indexing and integration on a Gaussian grid
was also performed. The total number of reflections observed was
17599, among which 2252 reflections were found to be independent and
1521 reflections were determined to satisfy the criterion
$I\,{>}\,2{\sigma}(I)$.  Structural parameters were refined by the
full-matrix least-squares method using SHELXL$-$97 software. The final
$R$ indices obtained for $I\,{>}\,2{\sigma}(I)$ were $R\,{=}\,0.0773$
and $wR\,{=}\,0.1638$. The crystal data are listed in Table
\ref{table:1}.  The chemical formula was confirmed to be
Ba$_2$CuSi$_2$O$_6$Cl$_2$ (BCuSOC).  The structure of BCuSOC is
orthorhombic $Cmca$ with cell dimensions of
$a\,{=}\,13.8917(12)$\,$\rm{\AA}$, $b\,{=}\,13.8563(11)$\,$\rm{\AA}$,
$c\,{=}\,19.6035(15)$\,$\rm{\AA}$ and $Z\,{=}\,16$. Its atomic
coordinates and equivalent isotropic displacement parameters are shown
in Table \ref{table:2}.

The structure of BCuSOC, closely related to that of
Ba$_2$CoSi$_2$O$_6$Cl$_2$~\cite{Tanaka}, is orthorhombic $Cmca$ with
cell dimensions of $a\,{=}\,13.8917(12)$\,$\rm{\AA}$,
$b\,{=}\,13.8563(11)$\,$\rm{\AA}$ and
$c\,{=}\,19.6035(15)$\,$\rm{\AA}$. The crystal structure viewed along
the $b$, $a$ and $c$ axes is illustrated in
Figs.~\ref{fig:Cu_structure}(a), (b) and (c), respectively. The
crystal structure comprises CuO$_4$Cl pyramids with a Cl$^-$ ion at
the apex.  Magnetic Cu$^{2+}$ with spin-1/2 is located approximately
at the center of the base composed of O$^{2-}$, which is parallel to
the $ab$ plane. Two neighboring CuO$_4$Cl pyramids along the $c$ axis
are placed with their bases facing each other. The CuO$_4$Cl pyramids
are linked via SiO$_4$ tetrahedra in the $ab$ plane, as shown in
Fig.~\ref{fig:Cu_structure}(c). The atomic linkage in the $ab$ plane
is approximately the same as that of
BaCuSi$_2$O$_6$~\cite{Finger,Sparta,Sasago}.  Two bases of neighboring
CuO$_4$Cl pyramids are rotated in opposite directions around the $c$
axis, as observed in BaCuSi$_2$O$_6$. Such rotation of the bases of
the CoO$_4$Cl pyramids is absent in
Ba$_2$CoSi$_2$O$_6$Cl$_2$~\cite{Tanaka}.

\section{Results and analyses}
%\subsection{Magnetic susceptibilities\label{sus}}
 
\begin{figure}[ht!]
\begin{center}
\includegraphics[width=8.0cm,clip]{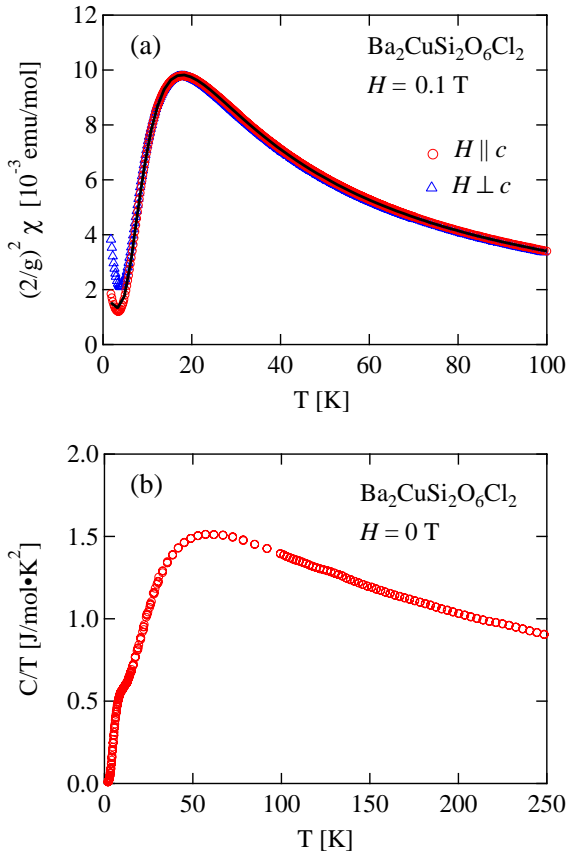}
\caption{(Color online) (a) Temperature dependence of magnetic 
  susceptibilities of BCuSOC measured for $H\,{\parallel}\,c$ and
  $H\,{\perp}\,c$, which are normalized by the $g$-factors of
  $g_{\parallel}\,{=}\,2.32$ and $g_{\perp}\,{=}\,2.06$. The solid line
  denotes the fit obtained using Eq.~(\ref{eq:kai}). (b) Temperature
  dependence of total specific heat divided by the temperature
  measured at zero magnetic field.}
\label{fig:sus_heat}
\end{center}
\end{figure}

Figure~\ref{fig:sus_heat}(a) shows the temperature dependence of the
magnetic susceptibilities ${\chi(T)}$ for BCuSOC with
$H\,{\parallel}\,c$ and $H\,{\perp}\,c$. The susceptibility data are
normalized by the $g$-factors of $g_{\parallel}\,{=}\,2.32$ and
$g_{\perp}\,{=}\,2.06$, determined from electron paramagnetic
resonance data at room temperature. The normalized susceptibilities
$(2/g)^2{\chi}(T)$ almost perfectly coincide above 6 K, which
indicates that the anisotropy of the magnetic interactions is very
small.  The susceptibility has a rounded maximum at $T_{\rm
  max}\,{\simeq}\,18$ K, followed by a rapid decrease with decreasing
temperature. No anomaly indicative of a phase transition was observed
down to the lowest temperature of $1.8$ K. The upturn in ${\chi(T)}$
below $4$ K is likely due to unpaired spins produced by lattice
defects. The overall feature of ${\chi(T)}$ is characteristic of
Heisenberg spin dimer magnets with a gapped singlet ground state.

\begin{figure}[ht]
\centering
\includegraphics[width=6.0cm,clip]{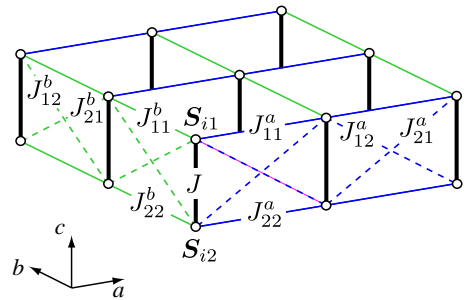}
\caption{(Color online) 2D model of the exchange network in 
  Ba$_2$CuSi$_2$O$_6$Cl$_2$. Thick solid lines represent the
  intradimer exchange interaction $J$, and thin solid and dashed lines
  respectively represent the interdimer exchange interactions
  $J^a_{{\alpha}{\beta}}$ and $J^b_{{\alpha}{\beta}}\ 
  ({\alpha},{\beta}=1, 2)$. }
\label{fig:exchange}
\end{figure}

It is natural to assume that the two Cu$^{2+}$ spins located on the
bases of neighboring CuO$_4$Cl pyramids along the $c$ axis are coupled
to form an AFM dimer. Because magnetic excitation in BaCuSi$_2$O$_6$
is dispersive in the $ab$ plane and dispersionless along the $c$
direction~\cite{Sasago}, BCuSOC can approximate a 2D coupled spin
dimer system with the exchange network shown in
Fig.~\ref{fig:exchange}. This is confirmed by the DFT calculation
shown later. Because the crystal structure is nearly tetragonal, we
can assume that
$J^a_{11}\,{=}\,J^a_{22}\,{=}\,J^b_{11}\,{=}\,J^b_{22}\,{\equiv}\,J_{\rm
  p}$ and
$J^a_{12}\,{=}\,J^a_{21}\,{=}\,J^b_{12}\,{=}\,J^b_{21}\,{\equiv}\,J_{\rm
  d}$.

When interdimer exchange interactions are treated as mean fields, the
magnetic susceptibility is expressed by
\begin{equation}
%{\chi}(T)={g^2{\mu_{\rm B}}^2{\beta}N}/\left({\exp{{\beta}J}+3+{\beta}{J^{\prime}}}\right),
{\chi}(T)=\frac{g^2{\mu_{\rm B}}^2{\beta}N}{\exp{({\beta}J)}+3+{\beta}{J^{\prime}}},
\label{eq:kai}
\end{equation}
where ${\beta}\,{=}\,1/(k_{\rm B}T)$, $N$ is the number of spins, $J$
is the intradimer exchange interaction and $J^{\prime}$ is the sum of
the interdimer interactions acting on one spin in a dimer, which is
given for the present system by $J^{\prime}\,{=}\,4(J_{\rm p}+J_{\rm
  d})$. Fitting Eq.~(\ref{eq:kai}) with the Curie and constant terms
to the experimental susceptibility for $H\,{\parallel}\,c$, we
evaluate $J$ and $J^{\prime}$ as $J/k_{\rm B}\,{=}\,29.4$ K and
$J^{\prime}/k_{\rm B}\,{=}\,13.8$ K, which are consistent with
$J/k_{\rm B}\,{=}\,28.1$ K and $J^{\prime}/k_{\rm B}\,{=}\,17.2$ K
evaluated from the analysis of the magnetization curve, shown below.
The solid line in Fig.~\ref{fig:sus_heat}(a) denotes the fit, which
excellently reproduces the experimental susceptibility.
%On the whole, these exchange parameters are consistent with those
%obtained from the analysis of the magnetization process shown below. 

\begin{figure}[ht!]
\begin{center}
\includegraphics[width=8.5cm,clip]{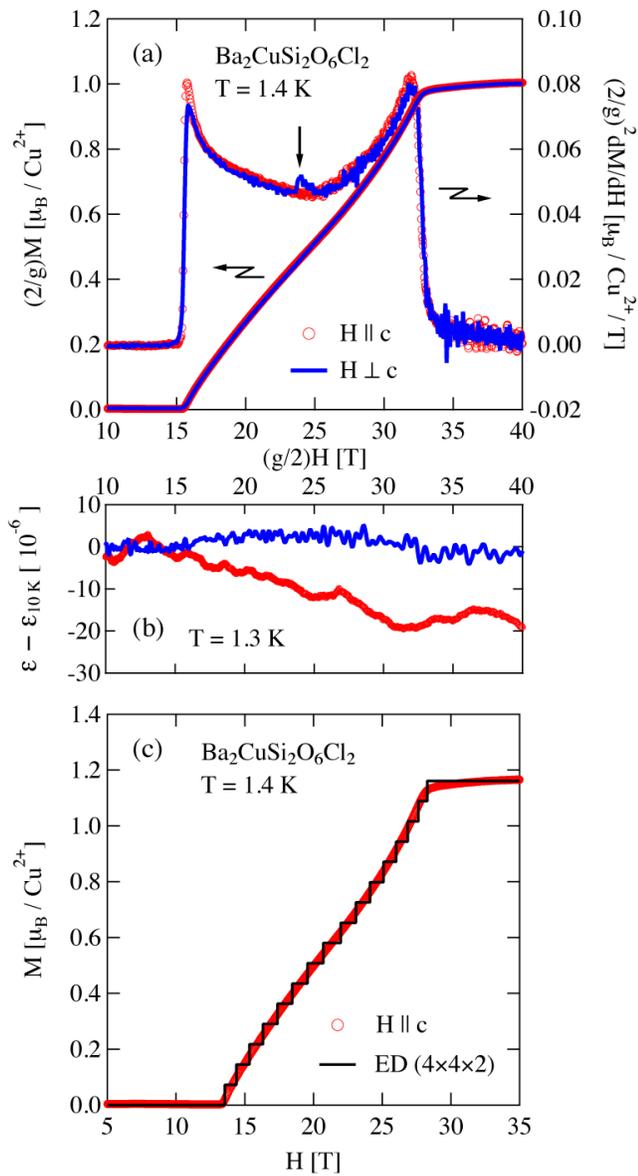}
\caption{(Color online) (a) Magnetic-field dependence of magnetization
  $M$ and its field derivative $dM/dH$ for BCuSOC measured at 1.4 K
  for $H\,{\parallel}\,c$ and $H\,{\perp}\,c$, which are normalized by
  the $g$-factors. An arrow indicates an anomaly in $dM/dH$ for
  $H\,{\perp}\,c$ owing to a phase transition accompanied with a small
  magnetization jump. (b) Axial magnetostriction $\varepsilon_a(H)$,
  with $H\,{\perp}\,c$ and $H\,{\parallel}\,c$, at $T\,{=}\,1.3$\,K
  after subtraction of the smooth background of similar magnitude
  measured at $T\,{=}\,10$\,K. The magnetostriction along and
  perpendicular to the $c$ axis (at $(g/2)H\,{=}\,25$\,T) show magnitudes and signs which, in a
  magnetically isotropic system, would indicate conservation of the
  unit cell volume. (c) Comparison between experimental and
  theoretical magnetization curves for $H\,{\parallel}\,c$.  The
  theoretical result was obtained using exact diagonalization (ED) for
  a 32-site spin cluster.}
\label{fig:MH}
\end{center}
\end{figure} 

Figure~\ref{fig:sus_heat}(b) shows the total specific heat divided by
the temperature of BCuSOC measured at zero magnetic field. No anomaly
indicative of a structural phase transition was observed down to 0.5
K, although in BaCuSi$_2$O$_6$, a structural phase transition was
observed at $T_{\rm s}\,{\simeq}\,100$ K~\cite{Samulon2}. In
BaCuSi$_2$O$_6$, there are three types of dimer with different
exchange interactions $J$ below $T_{\rm s}$~\cite{Rueegg4}. The
absence of the structural phase transition in BCuSOC indicates that
all the dimers are likely magnetically equivalent down to 0.5 K. A
Schottky-like anomaly is observed around 10 K owing to the excitation
gap of ${\Delta}/k_{\rm B}\,{=}\,20.8$ K, which is determined through
the high-magnetic-field magnetization measurement described below.

%\subsection{Magnetization process\label{MH}}

Figure~\ref{fig:MH}(a) shows the magnetization curves of BCuSOC and
their field derivatives measured at 1.4 K for $H\,{\parallel}\,c$ and
$H\,{\perp}\,c$ upon sweeping the magnetic field upward. The
magnetization data were normalized by the respective $g$-factors.  The
measurement was performed up to a magnetic field of 45 T using an
induction method with a multilayer pulse magnet at the Institute for
Solid State Physics, University of Tokyo. The entire magnetization
process was observed within the experimental field range.

The magnetization curves and their field derivatives for
$H\,{\parallel}\,c$ and $H\,{\perp}\,c$ almost perfectly coincide,
except for a small jump indicated by the vertical arrow in
Fig.~\ref{fig:MH}(a). These results confirm, as for ${\chi}(T)$
discussed before, that the anisotropy in the spin-spin interactions is
very small. The critical field $H_{\rm c}$ and saturation field
$H_{\rm s}$ are determined to be $(g/2)H_{\rm c}\,{=}\,15.5$\,T and
$(g/2)H_{\rm s}\,{=}\,32.8$\,T. From $H_{\rm c}$, the excitation gap
is evaluated to be ${\Delta}/k_{\rm B}\,{=}\,20.8$\,K. Note that
$H_{\rm s}/H_{\rm c}\,{=}\,2.12$ for BCuSOC is almost the same as
$H_{\rm s}/H_{\rm c}\,{=}\,2.1$ in BaCuSi$_2$O$_6$, where $H_{\rm
  c}\,{=}\,23.5$\,T and $H_{\rm s}\,{=}\,49$\,T for
$H\,{\parallel}\,c$~\cite{Jaime}.  For BCuSOC, the magnetization curve
for $H_{\rm c}\,{<}\,H\,{<}\,H_{\rm s}$ is ``inverse S" shaped, which
is characteristic of a low-dimensional quantum antiferromagnet.  The
linear magnetization slope observed in BaCuSi$_2$O$_6$~\cite{Jaime} is
likely a consequence of having three types of dimer.

The axial magnetostriction
$\varepsilon_a\,{=}\,(L(H)\,{-}\,L(0))/L(0)$ for $H\,{\parallel}\,c$
and $H\,{\perp}\,c$ was measured at 1.3\,K up to 45\,T and is shown in
Fig.~\ref{fig:MH}(b) after subtraction of the smooth background
obtained at $T\,{=}\,10$\,K. $\varepsilon_a(H)$ shows a lattice
response $\Delta L/L$ in the 10$^{-5}$ range for both orientations in the middle of
the field-induced magnetically ordered state in BCuSOC, i.\,e., at $(g/2)H\,{=}\,25$\,T. The observed
magnitude is rather small for an insulator and similar to or smaller
than values observed for other BEC systems where geometrical
frustration is weak. Indeed, the magnetostriction in the archetypical
system NiCl$_2$-4SC(NH$_2$)$_2$ was also found to be in the 10$^{-4}$
to 10$^{-5}$ range~\cite{weickert}. Here, the lattice response to an
applied field was used to quantify spin-spin correlations which, in
combination with the measured Young's modulus, were instrumental in
computing the dependence of the superexchange constant $J$ on the Ni
interionic distance $z$, $dJ/dz$. While similar computations for
BCuSOC are beyond the scope of this letter, it is a topic worth
exploring in future research~\cite{weickert}.

Because the spin dimers are parallel to the $c$ axis and the
magnetization shows no anomaly between $H_{\rm c}$ and $H_{\rm s}$ for
$H\,{\parallel}\,c$, we infer that $U(1)$ symmetry with respect to the
$c$ axis exists in BCuSOC. We analyzed the magnetization process for
$H\,{\parallel}\,c$ using exact diagonalization (ED) calculation for a
32-site ($4\,{\times}\,4\,{\times}\,2$) spin cluster on the basis of
the exchange model shown in Fig.~\ref{fig:Cu_structure}(d). Varying
the exchange parameters, we compared the calculated results with the
experimental magnetization curve. The best fit was obtained with
$J/k_{\rm B}\,{=}\,28.1$\,K, $J_{\rm p}/k_{\rm B}\,{=}\,3.9$\,K and
$J_{\rm d}/k_{\rm B}\,{=}\,0.4$\,K or with $J_{\rm p}/k_{\rm
  B}\,{=}\,0.4$\,K and $J_{\rm d}/k_{\rm B}\,{=}\,3.9$\,K. The solid
line in Fig.~\ref{fig:MH}(c) is the magnetization curve calculated
with these exchange parameters. The agreement between the experimental
and theoretical magnetization curves is excellent. The present results
demonstrate that BCuSOC closely approximates the 2D coupled Heisenberg
spin dimer system. Within the ED calculation, we cannot determine
whether the parallel $J_{\rm p}$ or diagonal $J_{\rm d}$ interdimer
interaction is dominant. This point will be addressed below using DFT calculations.

For $H\,{\perp}\,c$, $M(H)$ shows a small jump at
$M\,{\simeq}\,0.5M_{\rm s}$, which is evident in $dM/dH$
(Fig.~\ref{fig:MH}(a)). The magnetization anomaly observed upon
sweeping the magnetic field upward and downward is intrinsic.  It is
considered that the magnetization jump arises from spin reorientation.
Because the inversion symmetry with respect to the midpoint of the two
spins in a dimer is absent, the Dzyaloshinskii-Moriya (DM) interaction
of the form ${\bm D}_{i}\,{\cdot}\,[{\bm S}_{i1}\,{\times}\,{\bm
  S}_{i2}]$ between spins ${\bm S}_{i1}$ and ${\bm
  S}_{i2}$~\cite{Dzyaloshinsky,Moriya} can be finite, although it is
small. The $\bm D$ vector must be in the $ac$ plane because of the
twofold axis passing through the midpoint of the dimer along the $b$
axis~\cite{Moriya}. The magnetization curve for $H\,{\parallel}\,c$
shows no anomaly, hence the $a$ axis component of the ${\bm D}$ vector
will be small. The $c$ axis components of the ${\bm D}$ vectors on the
neighboring dimers should be antiparallel dueowing to the mirror plane
at $a/2$ and the glide planes at ${\pm}c/4$. The DM interaction is
usually accompanied with anisotropy of the form $({\bm
  S}_{i1}\,{\cdot}\,{\bm D}_{i})({\bm S}_{i2}\,{\cdot}\,{\bm
  D}_{i})/(2J)$ (known as the KSEA
interaction)~\cite{Kaplan,Shekhtman,Uchinokura}. The KSEA interaction
favors spins parallel to the ${\bm D}$ vector, which is assumed to be
parallel to the $c$ axis, while the DM interaction favors spins
perpendicular to the $c$ axis. Thus, competition between the DM and
KSEA interactions occurs, likely giving rise to a spin reorientation
transition for $H\,{\perp}\,c$.

\section{Density functional calculations}

\begin{table*}[ht]
  \caption{ \label{tab:dft} 
   Magnetic couplings in  Ba$_2$CuSi$_2$O$_6$Cl$_2$: Cu-Cu distances
   $d$ [\AA], hopping integrals $t$ [meV] 
  and magnetic couplings $J$ calculated 
  using GGA+$U_{\mathrm{scf}}$ [K]. Hopping integrals are 
  related to the AFM component 
  of the couplings through the relation 
  $J_{\alpha\beta}^{\mathrm{AFM}}\,{=}\,4t_{\alpha\beta}^2/U_{\mathrm{eff}}$ 
  where $U_{\mathrm{eff}}$ is the effective Hubbard repulsion term. 
  Positive couplings imply AFM interactions.}
   \begin{tabular}{L{1.5cm} C{1.2cm} | C{1.2cm} C{1.2cm}
   C{1.2cm}C{1.2cm} | C{1.2cm} C{1.2cm} C{1.2cm} C{1.2cm} | C{1.2cm} C{1.2cm}}
\hline
\hline
  &  $t$ & $t_{11}^{a}$ & $t_{22}^{a}$ & $t_{11}^{b}$ &  $t_{22}^{b}$ & $t_{12}^{a}$ & $t_{21}^{a}$ & $t_{12}^{b}$ & $t_{21}^{b}$ & $t_{\perp}$ & $t_{\perp}'$ \\
\hline    
$d$ (in~\AA) & 2.90 & 6.95 & 6.94 & 6.93 & 6.93  & 7.53 & 7.53 & 7.51 & 7.51 & 7.68 & 7.77 \\  
$t$ (in~meV)  & $-64$ & 5  & 8  & 5 & 5 &  28  & 28  &  32  & 32  &  0  &  0  \\  
\hline
\hline
           & $J$  & \multicolumn{4}{c|}{$J_{\rm p}$}  &  \multicolumn{4}{c|}{$J_{\rm d}$}   & \multicolumn{2}{c}{$J_{\perp}$} \\  
\hline
$J$ (in~K) & 30.5 & \multicolumn{4}{c|}{$-0.2$}     &  \multicolumn{4}{c|}{5.9}   & \multicolumn{2}{c}{$-0.3$} \\  
\hline
\hline
      \end{tabular}
\end{table*}

\begin{figure}[tb!]
\begin{center}
\includegraphics[scale=0.35]{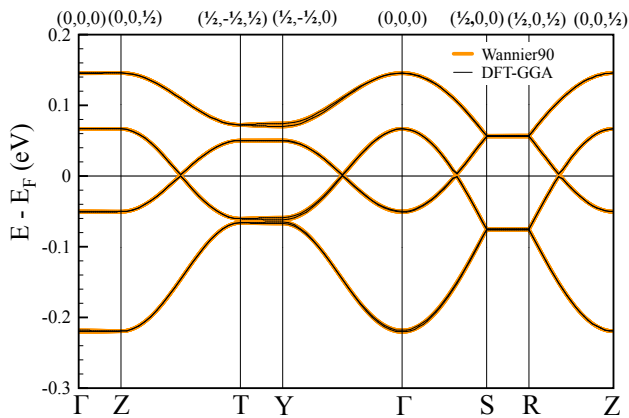}
\caption{ \label{fig:bands}(color online) 
Detail of the Cu-$3d_{x^2-y^2}$ band manifold around the Fermi level
in Ba$_2$CuSi$_2$O$_6$Cl$_2$ calculated using paramagnetic GGA-PBE and
interpolated with MLWFs. }
\end{center}
\end{figure} 

To assess the validity of our discussion on the basis of a microscopic
analysis, isotropic magnetic couplings were estimated using DFT
calculations~\cite{QuantumEspresso,Garrity14}.
As the first step, Wannier function interpolation~\cite{Wannier90} of
the GGA-PBE~\cite{Perdew96} paramagnetic bands of the dominant
Cu-$3d_{x^2-y^2}$ character was carried out. 
The results are shown in Figure~\ref{fig:bands}. The
corresponding hopping integrals are summarized in Table~\ref{tab:dft}.
An estimate of the AFM contribution to the magnetic couplings can
indeed be achieved through mapping of the paramagnetic band structure
onto a single-band Hubbard model at half filling, eventually reducing
to an AFM Heisenberg model in the strongly correlated limit. In this
approach, the effective interaction is given by
$J_{\alpha\beta}^{\mathrm{AFM}}\,{=}\,4t_{\alpha\beta}^2 /
U_{\mathrm{eff}}$.  
It can be readily seen from
  Table~\ref{tab:dft} that the largest hopping integral ($t$) occurs
  between two Cu$^{2+}$ ions located within the same structural dimer,
  revealing the potential presence of a leading nearest-neighbor
  AFM coupling. The next largest interactions
  ($t_{12}^a = t_{21}^a \approx t_{12}^b = t_{21}^b \equiv t_{d}$)
  couple two Cu$^{2+}$ ions belonging to one of the four neighboring
  dimers but located in adjacent planes of the same bilayer.
  The remaining interactions within the bilayer ($t_{11}^a = t_{22}^a
  \approx t_{11}^b \approx t_{22}^b \equiv t_{p} $) are an order of
  magnitude smaller, and inter-bilayer interactions ($t_{\perp}$
  and $t_{\perp}^{'}$) are vanishingly weak.  This hierarchy in the
  order of magnitude can easily be seen in Figure~\ref{fig:bands}
  using a simple 2D tight-binding analysis. The use of a 2D model is justified by the absence of
a sizable dispersion along $\Gamma \rightarrow Z$ or $S \rightarrow R$,
and therefore, the absence of sizable inter-bilayer interactions. 
Neglecting the inter-bilayer interactions
  $t_{\perp}$ and $t_{\perp}'$, the Cu-$3d_{x^2-y^2}$ band energy
  dispersions are given by
\begin{eqnarray}
\left.\begin{array}{l}
 \epsilon_{a,\pm}(\mathbf{k}) = \epsilon_d + t \pm 4 (t_{\rm p} + t_{\rm d}) \cos(\pi k_x) \cos(\pi k_y), \vspace{3mm}\\
 \epsilon_{b,\pm}(\mathbf{k}) = \epsilon_d - t \pm 4 (t_{\rm p} - t_{\rm d}) \cos(\pi k_x) \cos(\pi k_y),
\end{array}
\right\}\hspace{2mm}
\end{eqnarray}
 where $t_{\rm p}$ is the interdimer
 hopping parallel to the $ab$ plane, $t_{\rm d}$ is the diagonal interdimer hopping and ${\bm k}=(k_x, k_y)$ is the 2D reciprocal vector in units of the
 reciprocal lattice basis vectors. These two groups of bands are
 shifted with respect to each other by twice the intradimer
 hopping $t$, whose value can readily be determined at $S$, or to a
 lesser extent because of additional splittings occurring under the
 effect of further couplings at $T$ or $Y$. The band dispersion, as
 observed along $\Gamma \rightarrow S$ for instance, is directly
 related to the interdimer interaction $t_{\rm d}$ either through $(t_{\rm p}
 + t_{\rm d})$ for the lower-energy group of bands or through $(t_{\rm p} -
 t_{\rm d})$ for the higher-energy group.
 
 These results provide a clear picture of the dominant AFM magnetic
 couplings in Ba$_2$CuSi$_2$O$_6$Cl$_2$: the largest interaction
 $J^{\mathrm{AFM}}$ arises within the structural dimers, which are
 then coupled together by a weaker interdimer interaction
 $J_{\rm d}^{\mathrm{AFM}}$. Their ratio can be estimated to be
 $J_{\rm d}^{\mathrm{AFM}} / J^{\mathrm{AFM}}$ = $(t_{\rm d}/t)^2 \approx
 0.22$. Other interactions, arising either within the bilayers or
 between them, exhibit a negligible AFM component.
 
\begin{figure}[htb]
\begin{center}
  \includegraphics[scale=0.75]{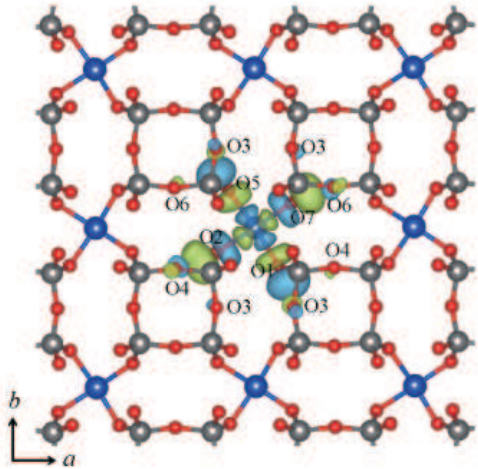}
\caption{ \label{fig:wannier} 
  (color online) MLWF centered on a Cu site. Large antibonding O-$2p$
  tails are clearly visible on the Cu coordinating oxygen atoms.}
\end{center}
\end{figure} 

Figure~\ref{fig:wannier} shows the MLWF corresponding to the band
interpolation presented in Fig.~\ref{fig:bands}. Its dominant
Cu-$3d_{x^2-y^2}$ character is clearly apparent, as well as the large
antibonding tails held by the four coordinating oxygen atoms O1, O2,
O5 and O7. Note however, that non-negligible components
are also visible on the oxygen atoms bridging two nonmagnetic
[SiO$_4$]$^{4-}$ units, i.e. O3, O4 and O6. As already discussed by
Mazurenko \textit{et al.} for BaCuSi$_2$O$_6$~\cite{Mazurenko14}, the
larger interdimer coupling $J_{\rm d}$ ($J_{12}\,{=}\,J_{21})$ occurs
between Cu$^{2+}$ ions belonging to adjacent layers and involves three
oxygen atoms. This essentially follows from the larger orbital overlap
along paths such as Cu$-$O5$-$O6$-$O7$-$Cu and Cu$-$O1$-$O4$-$O2$-$Cu along $a$ or
Cu$-$O2$-$O3$-$O4$-$Cu and Cu$-$O1$-$O3$-$O5$-$Cu along $b$ associated with larger
$\widehat{{\rm O}-{\rm O}-{\rm O}}$ angles ($\approx$~150\,$-$\,151$^{\circ}$) than
those involved in the same type of path for $J_{\rm p}$ ($J_{11}\,{\approx}\,
J_{22}$), i.e., between Cu$^{2+}$ ions of adjacent dimers belonging to
the same atomic layer (from $\approx$~87 to 141$^{\circ}$).

\begin{figure}[htb]
\begin{center}
   \includegraphics[scale=0.55]{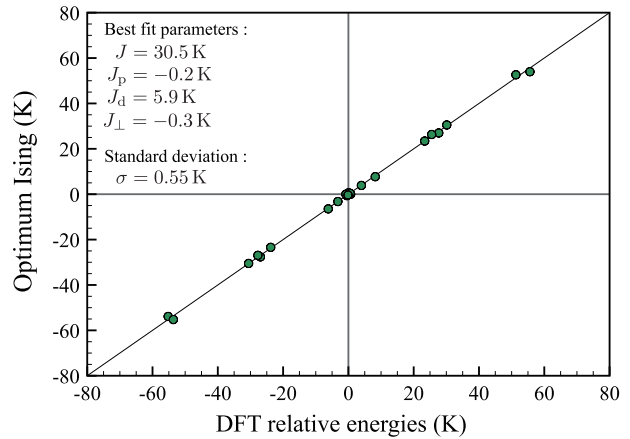}
\caption{ \label{fig:fit} 
(color online) Graphical representation of the
results obtained by the least-squares fitting procedure employed to
evaluate the magnetic couplings: optimized energies calculated
with the Ising Hamiltonian~(\ref{Ising}) are represented as a
function of the DFT + $U_{\mathrm{scf}}$ relative energies.
Positive couplings correspond to AFM interactions according to 
Eq. (\ref{Heisenberg}).}
\end{center}
\end{figure}

Alternatively, magnetic couplings were evaluated by energy
differences using the total energies calculated by GGA+$U_{\mathrm{scf}}$
for 22 distinct spin configurations obtained in the 104-atom
base-centred orthorhombic unit cell. DFT relative energies are shown
in Fig.~\ref{fig:fit} as a function of their optimal Ising counterparts
for the simplified model represented in Figure~\ref{fig:exchange}.
The excellent fit obtained for the entire set of spin
configurations further justifies the suitability of the simplified
Heisenberg Hamiltonian for modeling the magnetic properties of
Ba$_2$CuSi$_2$O$_6$Cl$_2$. Their values are presented in Table
\ref{tab:dft}. These calculations confirm that the dominant magnetic
coupling is AFM, $J/k_{\rm B}\,{=}\,30.5$~K, and that it occurs within structural dimers
coupled antiferromagnetically through a weaker interaction, $J_{\rm d}/k_{\rm B}\,{=}\,5.9$~K.  
Note that the amplitude ratio between the interdimer and intradimer
couplings, $J_{\rm d}/J \approx 0.19$, is in close agreement with the
value deduced from the analysis of the paramagnetic band structure (${\approx}\,0.22$) but slightly overestimates the ratio deduced from the analysis of the magnetization curves (${\approx}\,0.13$). The remaining couplings are very weak and ferromagnetic (FM). The
overall microscopic picture of the magnetism in
Ba$_2$CuSi$_2$O$_6$Cl$_2$ provided by density functional calculations
is therefore very similar to that of BaCuSi$_2$O$_6$~\cite{Mazurenko14}, where interdimer
coupling $J_{\rm d}$ connects spins belonging to distinct atomic planes
forming the bilayer.
Furthermore, the presence of weak FM couplings
between adjacent spins belonging to the same atomic plane or between
the bilayers implies the absence of obvious frustration, in contrast to the case of Ba$_2$CoSi$_2$O$_6$Cl$_2$~\cite{Tanaka}.  

\section{Conclusion}
In conclusion, we have presented the first structural analysis,
magnetization and magnetoelastic measurements on
Ba$_2$CuSi$_2$O$_6$Cl$_2$. The magnetic susceptibilities and
magnetization curves for different field directions coincide almost
perfectly when normalized by the respective $g$-factors, which
indicates that the magnetic anisotropy is very small. The observed
magnetization data are excellently reproduced using
exact-diagonalization computations based on the 2D coupled dimer model
illustrated in Fig.~\ref{fig:Cu_structure}(d). The 2D exchange model
was also confirmed by DFT.  Thus, Ba$_2$CuSi$_2$O$_6$Cl$_2$ closely
approximates the ideal 2D dimerized Heisenberg quantum magnet. The
quantum critical behavior near both critical fields $H_{\rm c,s}$ and
the nature of the spin state above $H_{\rm c}$ are of great interest
and currently under investigation.

\section*{Acknowledgment}
This work was supported by a Grant-in-Aid for Scientific Research (A)
(Grant No. 26247058), a Grant-in-Aid for Young Scientists (B) (Grant
No. 26800181) from Japan Society for the Promotion of Science, and the
German Research Foundation (Deutsche Forschungsgemeinschaft, DFG,
SFB-1143).  This work was performed using HPC resources from
GENCI-IDRIS (Grant No. 2016-i2016097218).  The NHMFL Pulsed Field
Facility is supported by the NSF, the U.S. DOE and the State of
Florida through NSF Cooperative Grant No. DMR-1157490.


\begin{thebibliography}{99}
\bibitem{footnote1} 
\bibinfo{note}{Note that, owing to the large (104 atoms) 
unit cell of Ba$_2$CuSi$_2$O$_6$Cl$_2$, 
the use of the supercell was not necessary in this case.}
\end{thebibliography}
\end{document}